\documentclass[prb,reprint,longbibliography]{revtex4-1}

\usepackage{amsmath}
\usepackage{amssymb}
\usepackage{graphicx}
\usepackage{xcolor}
\usepackage{hyperref}
\usepackage{bm}

\begin{document}

\title{Equation of motion approach to black-box quantization: taming the
multi-mode Jaynes-Cummings model}

\author{Fabian Hassler}
\author{Jakob Stubenrauch}
\author{Alessandro Ciani}

\affiliation{JARA-Institute for Quantum Information, RWTH Aachen University, D-52056 Aachen, Germany}

\begin{abstract} 
  An accurate modeling  of a Josephson junction that is embedded in an
  arbitrary environment is of crucial importance for qubit design.  We present
  a formalism to obtain a Lindblad master equation that describes the
  evolution of the system.  As the qubit degrees of freedom oscillate with a
  well-defined frequency $\omega_q$, the environment only has to be modeled
  close to this frequency.  Different from alternative approaches, we show
  that this goal can be achieved by modeling the environment with only few
  degrees of freedom. We treat the example of a transmon qubit coupled to a
  stripline resonator. We derive the parameters of a dissipative single-mode
  Jaynes-Cummings model starting from first principles. We show that the
  leading contribution of the  off-resonant modes is a correlated decay
  process involving both the qubit and the resonator mode. In particular, our
  results show that the effect of the off-resonant modes in the multi-mode
  Jaynes-Cummings model is perturbative in $1/\omega_q$.
\end{abstract}

\date{January 2019}

\maketitle

\section{Introduction}

The theory of open quantum systems is needed to describe measurements or
dissipation of a small quantum system such as a qubit. The most common
approach is to use a Lindblad master equation.\cite{nielsen} This equation is
local in time due to the fact that the density of states of the environment is
considered to be featureless and the coupling weak.\cite{breuer} The main
advantage of the Lindblad equation compared to more general
approaches\cite{leggett:87} is that it directly describes the evolution of the
reduced density matrix of the system without the need to solve for the
environmental degrees of freedom.  The study of systems that are coupled to
more realistic (linear) environments is of immediate relevance for a better understanding of quantum systems. In this case, the dynamics of the
environment is important and has to be treated appropriately.

For superconducting qubits, there has been recently a lot of interest in
investigating and understanding quite general environments. If the environment
is purely reactive,  an equivalent circuit consists only of inductances and
capacitances and can be quantized explicitly via one of the standard methods.
\cite{devoret:96,burkard:04,ulrich:16,ansari:18} More importantly, a rather
general approach called \emph{black-box quantization} has been put forward
recently for circuits with weak dissipation.\cite{nigg:12} A related method
that works for arbitrary strong dissipation that relies on results in
impedance synthesis has been proposed in
Refs.~\onlinecite{solgun:14,solgun:15}. Based on this method it was also shown
in Ref.~\onlinecite{solgun:17} that, considering the multi-port setting, the
parameters in the Hamiltonian can be fundamentally related to the elements of
the impedance matrix. All these methods have in common that a proper quantum
description of the environment involves many degrees of freedom.

These results do not follow the physical expectation that for a good qubit at
most a few degrees of freedom of the environment will be relevant.  The
physical intuition is even in contrast to recent findings that the multi-mode
Jaynes-Cummings model differs considerably from its single-mode
approximation.\cite{malekakhlagh:16,malekakhlagh:17, gely:17, parra:18} There
is thus a clear need for a \emph{general formalism} that allows to extract a
few relevant degrees of freedom of the environment and treats the rest in an
effective Lindbladian way.

\begin{figure}
  \centering
  \includegraphics[width=\linewidth]{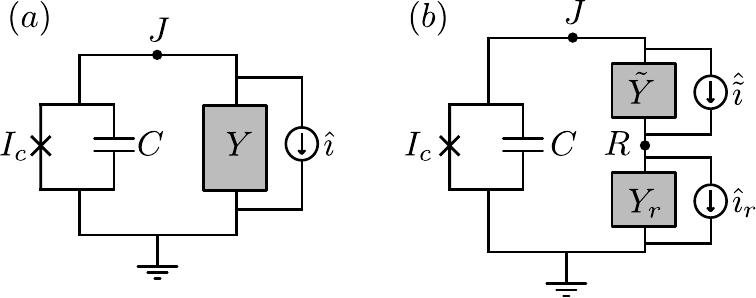}
  \caption{%
    ($a$) Setup of a qubit (consisting of a capacitance $C$ and a Josephson
    junction with critical current $I_c$) coupled to an arbitrary environment
    described by the admittance $Y_\omega$. In particular, we are interested
    in cases where the admittance is small and  is almost constant close to
    the frequency $\omega_q$ of the qubit. ($b$) If the latter condition is
    not fulfilled, we go over to an equivalent description of the system where
    the single admittance is replaced by two admittances $\tilde Y_\omega$ and
    $Y_{r,\omega}$ in series, such that $Y^{-1} = \tilde Y^{-1} +Y_r^{-1}$.
    The idea is that $Y_r$ is the admittance of a single bosonic mode and
    captures the frequency dependence whereas the (remaining) admittance
    $\tilde Y$ has a weaker frequency dependence.
  }\label{fig:setup}
\end{figure}

Here, we address this question: we present results for a superconducting qubit
that is embedded in a low admittance environment. We describe a
self-consistent procedure that decides if and how many modes have to be
extracted from the environment in order to obtain a good approximation of the
dynamics of the system. We show that due to the fact that in a qubit both the
voltage and the current fluctuate with the qubit frequency, the dynamics of
the environment only has to be accurately modeled close to this frequency.  In
particular, we discuss the case of a featureless environment and the case
where there is a single relevant degree of freedom.  We obtain explicit
expressions of the resulting Lindblad equation as a function of the admittance
of the environment. As an example, we treat a transmon qubit that is
capacitively coupled to a stripline resonator. We derive the effective
parameters of a dissipative Jaynes-Cummings model involving the qubit and a
single resonant mode of the cavity. We show that all the off-resonant modes
can be treated perturbatively. In particular, we find that the main effect of
the off-resonant modes is a correlated noise involving both the resonator and
the cavity.\footnote{We refer to the term proportional to $\gamma$ in
Eq.~\eqref{eq:lind_jc}. It is interesting that the leading effect of the
off-resonant modes is dissipative, see Eq.~\eqref{eq:deviation}.} Moreover, we
show that our formalism is capable of analytically describing the asymmetric
line-shape of the qubit decay rate that has been found in
Ref.~\onlinecite{houck:08}. 

The paper is organized as follows. In Sec.~\ref{sec::system} we introduce the
setup of a Josephson junction in parallel with an arbitrary admittance.
Considering the problem in the Heisenberg picture, we obtain the equation of
motion for the system operators. After projecting the equation of motion onto
the relevant qubit subspace and within the assumption that the environment
only weakly perturbs the qubit, we derive our central result, the approximate
equation of motion, Eq.~\eqref{eq:imp}, satisfied by the qubit. In
Sec.~\ref{sec::adm}, we introduce the admittance of a lossy stripline
resonator that serves as a concrete application for our general formalism
throughout the paper. The case in which the qubit is off-resonant with all the
modes of the environment (dispersive regime) is treated in
Sec.~\ref{sec:disp}.  In this case, the effect of the modes is just to cause a
shift of the qubit frequency as well as a decay which are connected to the
imaginary and real part of the admittance, respectively.  In
Sec.~\ref{sec::resReg}, we consider the case in which the qubit is close to a
resonance of one of the environmental modes. We explicitly show how the
resonant mode can be split off from the environment while still taking the
effect of the off-resonant modes into account. In particular, we obtain an
effective Jaynes-Cummings model where all the parameters are expressed in
terms of the admittance of the general environment.  We confirm our results
providing a comparison with numerical calculations in
Sec.~\ref{sec::numerics}. The conclusions are finally drawn in
Sec.~\ref{sec::conclusions}.

\section{System}
\label{sec::system}

We are interested in modeling a Josephson junction coupled to an arbitrary
environment. We denote the phase difference across the junction by
$\varphi(t)$ which is related to the voltage by the Josephson relation $V(t) =
(\hbar/2e) \dot \varphi(t)$ with $\hbar$ the reduced Planck's constant and
$e>0$ the elementary charge. We describe the influence of the linear
environment by an admittance $Y(t)$ that relates the voltage $V(t)$ to
the current $I_e(t)$ through the environment via
\begin{equation}\label{eq:admit}
  I_e(t) = \int_{-\infty}^\infty Y(t-t') V(t') \,dt' = (Y \star V)(t).
\end{equation}
Due to causality, its Fourier transform, given by $Y_\omega = \int e^{i\omega
t} Y(t)  dt$, is analytic with no poles or zeros in the upper half
plane.\cite{triverio:07} The real (imaginary) part of $Y_\omega$ describes the
dissipation (reactance).  \footnote{Note the different convention with respect
to the electrical engineering literature. In particular, the admittance of a
capacitance is given by $Y_\omega = -i \omega C$.} As the environment is
assumed to be linear the equation \eqref{eq:admit} is also the correct
relation between the current operator $\hat I_e$ and the voltage operator
$\hat V$ in the Heisenberg picture.

Kirchhoff's current law at the node $J$ in Fig.~\ref{fig:setup} demands that 
\begin{equation}\label{eq:ccl}
  \frac{\hbar C}{2 e} 
  \ddot {\hat\varphi}(t) + I_c \sin\hat{\varphi}(t) + \frac{\hbar}{2e} (Y \star
  \dot{\hat\varphi})(t) = \hat\imath(t)
\end{equation}
with $C$ ($I_c$) the capacitance  (critical current) of the Josephson junction
and $\hat \imath$ the noise due to the dissipative part of $Y$. 
Equation~\eqref{eq:ccl} is Heisenberg's equation of motion for the phase variable
$\hat\varphi(t)$. The noise is characterized by the commutation
relation\cite{zoller}
\begin{equation}\label{eq:comm}
  [\hat\imath(t), \hat\imath(t')] = i\hbar \frac{d}{dt}
  [Y(t-t')+Y(t'-t) ].
\end{equation}
Assuming that the dissipative elements of the environment (E) are
well-thermalized at a temperature $T$, the fluctuations are Gaussian random
variables with zero mean and a variance
\begin{equation}\label{eq:noise}
  \tfrac12 
  \langle \{\hat\imath^\dag_\omega, \hat\imath_{\omega'} \} \rangle_\text{E}
  =
  2 \pi \hbar \omega 
  \operatorname{Re} (Y_\omega)  (2 \bar n_\omega + 1)
  \delta(\omega -
  \omega')
  ,
\end{equation}
where $\{\hat A,\hat B\} = \hat A\hat B + \hat B \hat A$ denotes the
anticommutator and $\bar n_\omega = (e^{\hbar \omega/k_B T} -1)^{-1}$ is the
mean photon number.

Equation~\eqref{eq:ccl} is valid for a Josephson junction coupled to an
arbitrary linear environment. It is a true black-box equation, as the
environment only enters via its admittance and the associated noise term.
However, the equation is a non-linear stochastic operator equation for which
there are no clear solution strategies.  In the following, we will make a set
of controlled assumptions in which we extract a few relevant degrees of
freedom that evolve according to a Lindblad master equation. In particular, we
are interested in the situation where Eq.~\eqref{eq:ccl} describes the
dynamics of a qubit that is weakly perturbed by the environment which is
achieved in the small admittance setting. The specific requirements is given
in Eq.~\eqref{eq:cond_disp} and will be discussed below.

We first treat the environment as an open circuit ($Y=0$) and solve the
equation
\begin{equation}\label{eq:qubit_eom}
  \frac{\hbar C}{2 e} 
  \ddot {\hat\varphi}(t) + I_c \sin\hat{\varphi}(t) = 0;
\end{equation}
note that this is the Heisenberg equation associated with the  Hamiltonian
\begin{equation}\label{eq:hq}
  \hat H_q = - \frac{2 e^2}{C} \frac{\partial^2}{\partial \varphi^2} -
  \frac{\hbar I_c}{2e}
  \cos(\varphi).
\end{equation}
We assume that the dynamics only involves the two lowest eigenstates
$|g,e\rangle$ at frequencies $\omega_{g,e}$.\footnote{Of course, we have that
$\hat H_q |g,e \protect\rangle  = \hbar \omega_{g,e} |g,e \protect\rangle $.}
For the Hamiltonian $\hat H_q$ the two lowest eigenstates have opposite parity
with respect to $\hat\varphi \mapsto -\hat\varphi$ such that the phase
operator is off-diagonal in the eigenbasis.  As a result, at $Y=0$,
Eq.~\eqref{eq:qubit_eom} is solved by
\begin{equation}\label{eq:rw_ansatz}
  \hat \varphi(t) =
  \varphi_0
  \Bigl[ e^{-i\omega_q t} \hat\sigma^-(t) 
  + e^{i\omega_q t} \hat \sigma^+(t) \Bigr] .
\end{equation}
for constant $\hat \sigma^\pm(t) \equiv \hat\sigma^\pm$ with $\varphi_0 =
\langle e | \hat \varphi | g\rangle$ and $\omega_q = \omega_e - \omega_g$;
here and below, it will be convenient to parameterize the amplitude
$\varphi_0$ of the phase fluctuation defining the characteristic impedance of the
qubit $Z_q = \hbar \varphi_0^2/2 e^2$ and introduce the effective qubit
capacitance $C_q = (\omega_q Z_q)^{-1}$.\footnote{Note that in the transmon
limit the qubit capacitance $C_q$ coincides with the geometric capacitance
$C$.}

At weak dissipation, we can thus employ a variant of the rotating wave
approximation (RWA) in which we assume that $\hat\sigma^\pm_\eta$ contains
only frequency components with $|\eta| \ll \omega_q$. For concreteness, we
denote with $\bar\eta$ the typical frequency above which $\hat\sigma^\pm_\eta$
vanishes. In  Appendix~\ref{app:deri}, we show that with the ansatz
\eqref{eq:rw_ansatz} the projection of \eqref{eq:ccl} onto the relevant qubit
degrees of freedom leads to the equation of motion (valid for $\bar\eta \ll
\omega_q$)
\begin{align}\label{eq:imp}
  -2 (\hat \sigma^z \partial_t \hat \sigma^-)_\eta
   + \omega Z_q Y_{\omega} \hat\sigma^-_\eta
   &= i  \sqrt{\frac{2 Z_q}\hbar }  \hat \imath_\omega
\end{align}
in frequency space; here, $\eta = \omega- \omega_q$ is the frequency measured
with respect to the qubit frequency.   In the following, we will show how
Eq.~\eqref{eq:imp} can be used to analyze the influence of the environment on
the qubit in a few cases. In particular, the goal is to render Eq. \eqref{eq:imp}
into an equation that is local in time and thus can be related to a Lindblad
master equation.

In order that the approach taken above is valid, a few assumptions have to be
made about the environment. At first, we need that $\mathop{\rm Im} Y_\omega
\to 0$ for $\omega\to 0$. This excludes the case of a shunting inductance (as
in the fluxonium qubit) investigated in Ref.~\onlinecite{koch:09,smith:16}.
Moreover, as we will see below, we need that $|Y_{\omega_q}| Z_q \ll 1$. In
particular, this can be violated if there is an additional capacitance in the
environment.  In fact if $Z_q\mathop{\rm Im} Y_{\omega_q} \simeq 1$, we should
extract a capacitance from the environment and add it to $C$ before continuing
with the approach, see also below.

\section{Single- and multi-mode approximation of a stripline resonator}
\label{sec::adm}

In the previous section, we have obtained our central result
Eq.~\eqref{eq:imp}, which, under RWA, is valid for a general admittance.
In order to fix ideas, we connect  our general formalism to a concrete
physical setup of a transmon qubit that is capacitively coupled to a stripline
resonator whose admittance is introduced in this section. We will find a more
and more refined effective Liouvillian description in the reminder of the
paper. In particular, we want to show how all the modes of the stripline
resonator (forming the environment) can be naturally incorporated and that
there is a perturbative procedure to go beyond the single-mode approximation.
For low dissipation, the admittance of the resonator is approximately given by
\begin{equation}\label{eq:mm} Y_\text{mm} = -\frac{ 2i C_r \omega_0}{\pi} 
  \tan\left[\frac{\pi (\omega +
  \tfrac{i}2 \kappa)}{\omega_0} \right]
\end{equation}
with $\omega_0$ the fundamental frequency of the resonator,  $\kappa$ the
damping rate, and $C_r$ the characteristic capacitance of the resonator; see
App.~\ref{app:admit}. The modes of the resonator are at  $\omega_n = n
\omega_0$ with $n\in\mathbb{N}$.  Including the coupling capacitance, the
total impedance of the environment is given by $Y^{-1} = Y^{-1}_{c} +
Y^{-1}_\text{mm}$ with $Y_c = - i\omega C_c$. As we will see in the following,
the coupling capacitance $C_c$ (assumed to be small) is proportional to the
coupling rate $g$ of the Jaynes-Cummings Hamiltonian.

If we assume that the qubit frequency is close to the frequency of mode $n_0$,
i.e., $|\Delta| < \omega_0$ with the detuning $\Delta = \omega_q -
\omega_{n_0}$, we can approximate the admittance by a single mode ($\omega>0$)
\begin{align}\label{eq:Yr}
  Y_r 
  &= - 2i C_r (\omega -\omega_r +\tfrac12 i \kappa)
\end{align}
%
%
with $\omega_r = \omega_{n_0}$. This corresponds to the positive frequency
response of an RLC circuit with a capacitance $C_r$, a resistance $R_r = (C_r
\kappa)^{-1}$, and an inductance $L_r = (C_r \omega_r^2)^{-1}$ in parallel.
Note that close to the resonance with $\kappa , |\omega -\omega_r|  \ll
\omega_0$, the single-mode approximation is a very good approximation to the
total admittance.  In particular, one can model the stripline as the single
mode $Y_r$ in series with another admittance of value $ \tilde Y = (Y^{-1} -
Y^{-1}_r)^{-1}$. For small coupling ($C_c \ll C_r$), we find
the expansion
\begin{align}\label{eq:deviation}
  \tilde Y = (Y^{-1} - Y^{-1}_r)^{-1}
  &\approx Y_c + Y_c^2 (Y_r^{-1} - Y_\text{mm}^{-1})  \nonumber\\
  &\approx
  -i\omega C_c + \frac{ \pi^2 \omega^2 C_c^2 \kappa}{12 \omega_0^2 C_r}
\end{align}
valid close to the resonance frequency; the first term describes the influence
of the coupling capacitance whereas the second term captures the
leading contribution of the influence of \emph{all} the  modes that are
off-resonant. Its effect is dissipative and will lead to the decay constant
$\gamma$ in the Jaynes-Cummings model.

\section{Dispersive regime}\label{sec:disp}

Having set the stage by presenting the central result as well as our concrete
physical application, we will show how to derive a Lindblad master equation
for the case when the qubit is detuned from all the resonances of the
environment, i.e., we treat the case of a general though off-resonant
environment. In this case, we can assume that the admittance is constant over
the relevant frequency range $[\omega_q -\bar\eta, \omega_q + \bar\eta ]$. As
a result (provided that $\bar\eta \ll \omega_q$), Eq.~\eqref{eq:imp} becomes
local in time and assumes the simple form
%
%
\begin{equation}\label{eq:disp}
  \partial_t \hat \sigma^-(t)
  + \tfrac12 \omega_q Z_q Y_{\omega_q} \hat\sigma^-(t)
   =  \sqrt{\omega_q  Z_q
   \operatorname{Re}(Y_{\omega_q})  } \, \hat \sigma^z(t)\hat b(t)
\end{equation}
of a quantum Stratonovich  stochastic differential equation;\footnote{See
Sec.~3.4.5 of Ref.~\onlinecite{zoller}.} here, we have introduced  a new
operator  $\hat b(t)$ via
\begin{equation} \hat \imath(t) = i  \sqrt{2\hbar \omega_q
  \operatorname{Re}(Y_{\omega_q}) } e^{-i \omega_q t}
 \hat b(t) 
\end{equation}
The operator $\hat b(t)$ is a quantum white noise with  $[\hat b(t), \hat b^\dag(t')] =\delta(t-t')$, $[\hat b(t), \hat
b(t')] =0$,
\begin{equation}
  \langle \{\hat b(t), \hat b^\dag(t')\} \rangle_\text{E} = (2 \bar
  n + 1) 
    \delta(t-t') ,
\end{equation}
and $\langle \{\hat b(t), \hat b(t')\} \rangle_\text{E} = 0$ with $\bar n = \bar
n_{\omega_q}$.

By inspecting the left hand side of Eq.~\eqref{eq:disp}, we immediately observe
that the environment leads to a frequency shift (also called Lamb-shift)
\begin{equation}
\delta \omega_q = \tfrac12 \omega_q Z_q \operatorname{Im}(Y_{\omega_q})
\end{equation}
and a decay rate  (also called Purcell rate)
\begin{equation}
  \gamma = \omega_q Z_q \operatorname{Re}(Y_{\omega_q});
\end{equation}
see also Ref.~\onlinecite{esteve:86,houck:08,bronn:15,scheer:18}. As the
equation of motion (including the noise) is local in time, it is equivalent to
the Lindblad equation\footnote{Our derivation of the Lindblad equation does
not include the ac-Stark shift which is typically unimportant. A more refined
calculation leads to \eqref{eq:linblad_disp} with $(2\bar n +1)$ as the
prefactor of the first term proportional to
$\delta\omega_q$.\cite{carmichael1}\protect\vphantom.}
\begin{multline}\label{eq:linblad_disp}
  \dot \rho = - \frac{i}{2} \delta\omega_q [\hat \sigma_z, \rho] +
  \gamma (\bar n + 1)
  \mathcal{D}[\hat \sigma^-](\rho)  \\
  +   \gamma \bar n 
  \mathcal{D}[\hat \sigma^+](\rho) 
\end{multline}
for the density matrix of the qubit; here, we have introduced the
superoperator $\mathcal{D}[\hat J](\rho) = \hat J \rho \hat J^\dag - \tfrac12
\{\hat J^\dag \hat J, \rho\}$ corresponding to the jump operator $\hat J$. Now, we can formulate (self-consistently) when the dispersive approximation is
applicable. Indeed, we have that $\bar \eta \approx \max\{ |\delta \omega_q|
,\gamma\}$. As a result, the approach is valid as long as the admittance does
not change appreciably on the scale $\bar \eta$; i.e.,
Eq.~\eqref{eq:linblad_disp} is valid as long as the self-consistency equation
\begin{equation}\label{eq:cond_disp}
  \frac{d Y_{\omega_q}}{d \omega } \, \max\{ |\delta \omega_q|
,\gamma\} \ll Y_{\omega_q}
\end{equation}
is fulfilled.

For the concrete example of stripline resonator, introduced in
Sec.~\ref{sec::adm}, we can use the expansion
\begin{equation}\label{eq:disp_imp} 
  Y = Y_{c} - \frac{Y_c^2}{Y_\text{mm}}
\end{equation}
valid for weak coupling  with $|Y_c/Y_\text{mm} | \approx C_c/C_r \ll 1$. The
first term in \eqref{eq:disp_imp} corresponds to a capacitance in
\emph{parallel} to the junction capacitance. Its effect can be taken (exactly)
into account by replacing $C_q \mapsto C_q + C_c$.\footnote{In general, one
  can always include the effect of a capacitance of value $ i (d
Y/d\omega)(\omega=0)$ exactly and only treat the remaining admittance as the
environment.}  The nontrivial effects of the environment solely arise from the
second term. The admittance changes on the scale $\Delta$ such that for
self-consistency, we have to require that $ |\delta \omega_q| ,\gamma \ll
|\Delta|$.  For the stripline resonator, we obtain the expressions
\begin{align}\label{eq:disp_res}
  \delta \omega_q =  \frac{\pi g^2 \sin(2\pi \Delta/\omega_0)}{\omega_0
    [\cosh(\pi \kappa/\omega_0)-
  \cos(2\pi \Delta/\omega_0)] } ,\\\label{eq:disp_res_g}
  \gamma =  \frac{2 \pi g^2 \sinh(\pi \kappa/\omega_0)}{\omega_0
    [\cosh(\pi \kappa/\omega_0)-
  \cos(2\pi \Delta/\omega_0)] }; 
\end{align}
here, we have introduced the coupling rate
\begin{equation}\label{eq:g}
  g =  \frac{C_c}{2\sqrt{C_q C_r}} \omega_q
\end{equation}
which will later be shown to be the coupling constant of the Jaynes-Cummings
model. Note that we have consciously defined the coupling rate different from
the more common choice, with $\omega_q$ replaced by the symmetric expression
$\sqrt{\omega_r \omega_q}$. The reason is twofold. First, we consider an
initial situation where the qubit is excited while the resonator is still in
its equilibrium state. Due to this, it is more natural to evaluate the
environment at the qubit frequency. Second, due to this definition, the simple
expressions in \eqref{eq:disp_res_appr} and \eqref{eq:purcell} are valid all
the way up to order $(\Delta,\kappa)^4/\omega_0^4$ which is not true for the
alternative definition.

In the expressions \eqref{eq:disp_res} and \eqref{eq:disp_res_g} still all the
modes of the stripline resonator have been included. The only assumptions so
far are weak coupling ($C_c  \ll C_r$), small shifts ($ |\delta \omega_q|,
\gamma \ll |\Delta|$), and weak damping ($\kappa \ll \omega_0$).  For the case
of small detuning ($\Delta \ll \omega_0$), the expressions \eqref{eq:disp_res}
and \eqref{eq:disp_res_g} can be further simplified to
\begin{align}\label{eq:disp_res_appr}
  \delta \omega_q &=  \frac{g^2 \Delta}{\Delta^2 + (\kappa/2)^2} - \frac{\pi^2
  g^2 \Delta}{3 \omega_0^2} ,
   \\\label{eq:purcell}
   \gamma &=  \frac{g^2 \kappa}{\Delta^2 + (\kappa/2)^2}  + \frac{\pi^2 g^2 \kappa}{3 \omega_0^2} 
\end{align}
with a correction term only appearing in order $1/\omega_0^4$. Note that the
single-mode approximation corresponds to the first terms and the effect of all
the other modes is captured perturbatively by the second term.  The present
approach gives a perturbative  expansion in the small parameters and avoids
the rediagonalization of the complete system as in Ref.~\onlinecite{gely:17}.
In this expansion, the conventional single-mode approximation corresponds to
the first term,\cite{blais:04} with all the remaining modes contributing to a
small correction of order $\omega_0^{-2}$.

\section{Resonant regime}\label{sec::resReg}

In the previous section, we have treated the simplest case of a dispersive
qubit-environment coupling. In this case, the admittance does not vary too
much close to the qubit frequency and we can simply replace it by a constant.
A more elaborate analysis is necessary in the case where the shift is so large
that the admittance cannot be assumed to be constant and
Eq.~\eqref{eq:cond_disp} is violated. In particular, in the case of a small
admittance, this can only happen when there is a root of the admittance (in
the complex plane at $\omega_*$) close to the frequency of the qubit. Let us
parameterize the root in question by $\omega_*= \omega_r - i \kappa/2$ and
introduce the characteristic capacitance $C_r= \frac{i}2
(dY/d\omega)_{\omega=\omega_*}$.\footnote{Note that for weak dissipation the
capacitance is always real.} The admittance close to the qubit frequency is
then well approximated by a single mode $Y_r$.  We assume that the remaining
admittance $\tilde Y = (Y^{-1} -Y^{-1}_r)^{-1}$ does not change appreciably
over the range $\bar \eta$, i.e., that \eqref{eq:cond_disp} is satisfied for
$\tilde Y$ (even though it does not hold for $Y$ itself). Otherwise, the
process of extracting a mode can be repeated  until this assumption is
fulfilled.

The resonant mode corresponds to an RLC circuit. We associate with this
circuit a bosonic mode $\hat a$. Moreover, we  extend the equation of motion
to incorporate this mode. In particular, we have the equations of motion (for
the nodes $J$ and $R$ in Fig.~\ref{fig:setup})
\begin{align}\label{eq:imp_r}
  &- 2 (\hat 
  \sigma^z \partial_t \hat \sigma^-)_\eta + \omega Z_q \tilde Y_\omega \biggl(\hat
  \sigma^-_\eta -  \sqrt{\frac{C_q}{C_r}}\, 
\hat a_\eta\biggr) = i \sqrt{\frac{2 Z_q}{\hbar}}\,
\hat{\tilde \imath}_\omega\;, \\   
  &-2i C_r (\eta+\Delta+\tfrac{i}2 \kappa) \hat a_\eta + \tilde Y_\omega 
  \biggl(\hat a_\eta - \sqrt{\frac{C_r}{C_q}} \,\hat
  \sigma^-_\eta\biggr)\nonumber\\\label{eq:imp_r2}
  &\qquad\qquad\qquad\qquad\qquad\qquad
  = i \sqrt{\frac{2 C_r}{\hbar \omega_q}} ( \hat{
  \imath}_{r,\omega} - \hat{\tilde
\imath}_\omega )
\end{align}
where $\hat{\tilde\imath}$ ($\hat{\imath}_r$) corresponds to the noise due to
$\operatorname{Re}\tilde Y$ ($\operatorname{Re} Y_r$). It can be checked by a
straightforward calculation that solving \eqref{eq:imp_r2} for $\hat a_\eta$
and inserting the resulting expression into \eqref{eq:imp_r} that
\eqref{eq:imp} is recovered, see App.~\ref{app:int_out}. The advantage of the
new representation is that the frequency dependence of $\tilde Y$ around the
qubit frequency is milder than the one of $Y$. In particular, provided that
$\tilde Y$ does not change appreciably on the scale $\bar\eta$, we can replace
$\omega$ by $\omega_q$ and arrive at the system of equations
\begin{align}\label{eq:sys_eq_loc}
  \partial_t \hat \sigma^- (t) +  \frac{\tilde Y_{\omega_q}}{2 C_q}
  &\biggl[\hat
    \sigma^-(t) +  \sqrt{\frac{C_q}{C_r}} \hat \sigma^z(t) \hat a(t)\biggl] \nonumber\\
  &= \sqrt{C_q^{-1} \operatorname{Re} \tilde
  Y_{\omega_q}} \,
  \hat\sigma^z(t) \hat{\tilde b}(t) ,
   \\\label{eq:sys_eq_loc2}
  (\partial_t- i \Delta + \tfrac{1}2 \kappa) \hat a(t) &+\frac{ \tilde
  Y_{\omega_q}}{2 C_r}
  \biggl[\hat a(t) -  \sqrt{\frac{C_r}{C}}  \hat \sigma^-(t)\biggr]\nonumber\\
  & =
  \sqrt{C_r^{-1} \operatorname{Re} \tilde
  Y_{\omega_q}} \,
  \hat{\tilde b}(t)  - \sqrt{\kappa} \,\hat
  b_r(t) 
\end{align}
that are local in time.

The coherent evolution is generated by the Jaynes-Cummings Hamiltonian
\begin{equation}
  \hat H_\text{JC} =  \tfrac12 \delta \omega_q \hat \sigma_z + 
  (\delta \omega_r- \Delta) \hat a^\dag \hat a + g (\hat\sigma^+ \hat a + \hat a^\dag
  \hat \sigma^-),
\end{equation}
with the parameters
\begin{align}
  \delta \omega_q =   \frac{\operatorname{Im}(\tilde Y_{\omega_q})}{2 C_q}, \;
  \delta \omega_r =   \frac{\operatorname{Im}(\tilde Y_{\omega_q})}{2 C_r}, \;
  g = -\frac{\operatorname{Im}(\tilde Y_{\omega_q})}{2 \sqrt{C_q C_r}},
\end{align}
corresponding to the frequency shift of the qubit, the frequency shift of the
resonator, and the coherent coupling rate. Transforming the quantum
Stratonovich stochastic differential equations \eqref{eq:sys_eq_loc} and
\eqref{eq:sys_eq_loc2} to an equivalent Lindblad equation  yields
\begin{align}\label{eq:lind_jc}
  \dot \rho = - i [\hat H_\text{JC}, \rho] &+ \kappa (\bar n + 1)
  \mathcal{D}[\hat a](\rho)  +  \kappa \bar n
  \mathcal{D}[\hat a^\dag ](\rho) \nonumber\\& +  \gamma (\bar n + 1)
  \mathcal{D}\Bigl[\hat \sigma^- - \sqrt{\frac{C_q}{C_r}}\hat a\Bigr](\rho)
  \nonumber\\
 &+  \gamma \bar n
  \mathcal{D}\Bigl[\hat \sigma^+ - \sqrt{\frac{C_q}{C_r}}\hat a^\dag\Bigr](\rho) 
\end{align}
with the decay rate
\begin{align}
  \gamma&=   \frac{\operatorname{Re}(\tilde Y_{\omega_q})}{C_q}. 
\end{align}
This rate corresponds to a correlated decay process involving both the qubit
as well as the oscillator. The master equation is valid as long as the
self-consistency equation \eqref{eq:cond_disp} is fulfilled where $Y$ is
replaced by $\tilde Y$.

The Eq.~\eqref{eq:lind_jc} is the master equation of a Josephson junction
coupled to a general environment. For the concrete example of a stripline
resonator, we can use the expression \eqref{eq:deviation} when the qubit is
brought close to the $n_0$-th resonance of the resonator at the frequency
$\omega_r$. In this case, the parameters of the Jaynes-Cummings model assume
the form
\begin{align}
  \delta \omega_q &= - \frac{C_c \omega_q}{2 C_q}, & \delta \omega_r&= -
  \frac{C_c \omega_q}{2 C_r}, & g&= 
  \frac{C_c \omega_q}{2 \sqrt{C_q C_r}}.
\end{align}
The qubit frequency shift and the coherent coupling corresponds to the first
term expansion in $C_c$ of the results presented in Sec.~\ref{sec:disp}.
The leading contribution of the off-resonant modes is given by the correlated
decay with a rate
\begin{equation}
  \gamma = \frac{\pi^2 g^2 \kappa}{3\omega_0^2};
\end{equation}
this is the leading effect of the off-resonant modes in the multi-mode
resonator cf.\ Eq.~\eqref{eq:purcell}; to the best of our knowledge, the
correlated decay has never been discussed before in the literature.

\section{Comparison to numerics}\label{sec::numerics}

In this section, we would like to compare the results of Sec.~\ref{sec:disp}
to the numerical approaches of Refs.~\onlinecite{houck:08,gely:17}.
The numerical approaches are designed to work in the transmon regime with $I_c
\gg e^3/\hbar C$. In this regime, the Josephson junction acts as an inductance
$L_J = 2e/\hbar I_c$ as we  can use the approximate equality $I_c \sin \phi
\approx I_c \phi$ in Eq.~\eqref{eq:ccl}. With this, the total system becomes
linear and thus several solution strategies are available.

\begin{figure}
  \centering
  \includegraphics{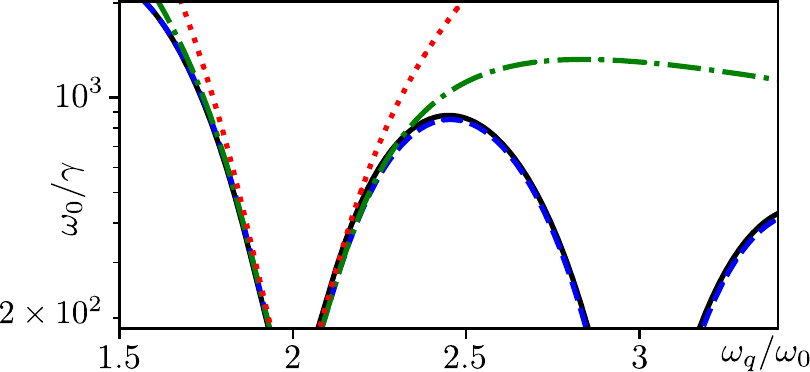}
  \caption{%
  Dimensionless relaxation time $\omega_0/\gamma^{-1}$ of a qubit due to the
  multi-mode environment with $C_c/C_r=0.02$, $C_c/C=0.02$, and
  $\kappa/\omega_0=0.2$ as a function of the qubit frequency $\omega_q$. The
  solid line is the numerics given by the \emph{admittance approach} of
  Ref.~\onlinecite{houck:08}. The plot shows $\gamma=
  -2\operatorname{Im}(\omega_*)$. The dashed line is the result of
  Eq.~\eqref{eq:disp_res_g}. The dash-dotted line is the approximate
  expression \eqref{eq:purcell} with $\omega_r = 2 \omega_0$. The dotted line
  is the single-mode approximation (first term of \eqref{eq:purcell}). It can
  be seen that the analytical formulas describe the decay rate rather well. In
  particular \eqref{eq:purcell} reproduces the asymmetry in the decay rate
  that has been observed in Ref.~\onlinecite{houck:08}.
  }\label{fig:purcell}
\end{figure}

The \emph{admittance approach} of Ref.~\onlinecite{houck:08}, directly solves
for the eigenmodes of the complete system by seeking the roots of the total
admittance
\begin{equation}
 Y_C + Y_J  + Y = 0
\end{equation}
as a function of $\omega$; here, $Y_C = -i \omega C$ and $Y_L =(-i \omega
L_J)^{-1}$ are the admittances of the capacitance and the linearized Josephson
junction respectively. The approach works for reactive as well as for
dissipative environments. For each solution $\omega_*$, the real part denotes
the frequency and the imaginary part the decay rate. As long as the admittance
of the environment is small, there is a solution close to the bare qubit
frequency which is plotted in Fig.~\ref{fig:purcell}. This approach is in
principle very close to our presentation in Sec.~\ref{sec:disp}. However,
crucially our approach works also outside the transmon regime where the total
system is not approximately harmonic. In Fig.~\ref{fig:purcell}, it can be
seen that the analytical results   Eqs.~\eqref{eq:disp_res} and
\eqref{eq:disp_res_appr} describe the behavior rather well. In particular, the
simple expression \eqref{eq:disp_res_appr}  captures the asymmetry of the
decay rate due to the higher modes that has been observed in
Ref.~\onlinecite{houck:08}.

\begin{figure}
  \centering
  \includegraphics{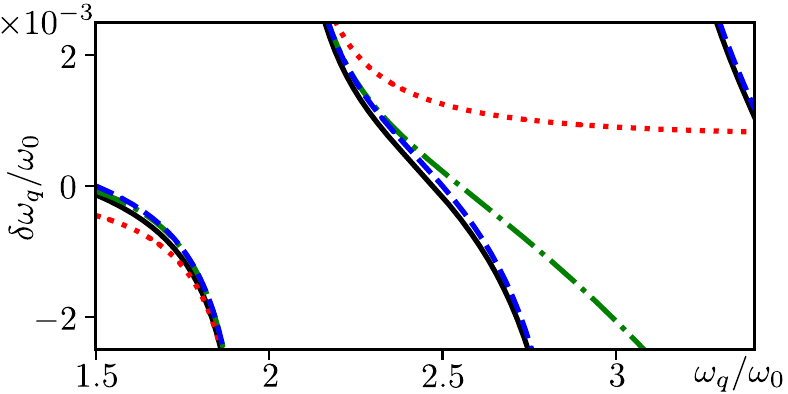}
  \caption{%
    Lamb shift: The parameters are chosen as in
    Fig.~\ref{fig:purcell} with $\kappa =0$. The solid line shows the
    numerical result of the \emph{rediagonlization approach} of
    Ref.~\onlinecite{gely:17}, i.e., Eq.~\eqref{eq:all_modes}. Note
    that our analytical results, Eq.~\eqref{eq:disp_res} (dashed line) and
    Eq.~\eqref{eq:disp_res_appr} (dash-dotted line), reproduce the numerical
    results rather well. Even the single-mode approximation, first term
    \eqref{eq:disp_res_appr}  (dotted line), is valid close to the resonance
    which indicates that Eq.~\eqref{eq:all_modes} is approximately given by the
    na\"ive result $g^2/\Delta$.
  }\label{fig:lamb_shift}
\end{figure}

The \emph{rediagonalization approach} of Ref.~\onlinecite{gely:17}
on the other hand, concentrates on reactive environments. It proceeds by
finding new eigenmodes of the stripline resonator (=environment) once they
are coupled to the qubit. For completeness, we show the procedure in
App.~\ref{app:rediag}. The result are a set of frequencies $\nu_n$ and
coupling constants $g_n$ which allow for treating the system as a multi-mode
Jaynes-Cummings model. Note that we do not have $\nu_n = n\omega_0$ anymore.
The Lamb shift then gets a contribution of all the eigenmodes which act
independently. In particular, we find
\begin{equation}\label{eq:all_modes}
  \delta \omega_q = \sum_{n=0}^\infty 
  \frac{g_n^2}{\Delta_n(1-\Delta_n/2\omega_q)}
\end{equation}
with $\Delta_n = \omega_q - \nu_n$. In Fig.~\ref{fig:lamb_shift}, we show that
there is an (approximate) sum rule associated with the new eigenmodes. In
particular, the contribution of all the modes (Eq.~\eqref{eq:all_modes})
approximately gives the result $g^2/\Delta$ of the bare single mode
Jaynes-Cummings model.  Moreover, we find that the contribution of the remaining
modes is well captured by the second term in Eq.~\eqref{eq:disp_res_appr}.

\section{Conclusion} \label{sec::conclusions}

In conclusion, we have presented a general approach to derive an effective
Lindblad equation for a superconducting qubit embedded in an arbitrary linear
environment with a small admittance. The approach yields an effective model
where the qubit is coupled to a few  harmonic degrees of freedom of the
environment. For the case of a single relevant environmental mode, we have
given explicit expressions of the model parameters (coupling constant, decay
rates, jump operators) as a function of the admittance. In particular, we have
found that the single-mode Jaynes-Cummings model is the leading term in the
description of a superconducting qubit that is coupled to a multi-mode
resonator. The main effect of the off-resonant modes is a novel, correlated
decay that involves both the resonator and the qubit degrees of freedom. In
particular, our results show that the multi-mode Jaynes-Cummings model does
not only have a cutoff free description but that the correction of the
off-resonant modes can be obtained analytically in a perturbative expansion in
$1/\omega_0 \simeq 1/\omega_q$.  It is an interesting idea for further studies
to extend our results to the case of a fluxonium qubit. In this case, the
environment has a large admittance with $Z_q |Y_\omega| \gg 1$ such that an
perturbative approach using the impedance $Z_\omega=1/Y_\omega$ instead of the
admittance seems to be the appropriate as a starting point.

\begin{acknowledgments}
FH and AC acknowledge financial support from the Excellence Initiative of the Deutsche Forschungsgemeinschaft.
\end{acknowledgments}

\appendix

\section{Derivation of Eq.~\eqref{eq:imp}}\label{app:deri}

In this Appendix, we give a more rigorous derivation of Eq.~\eqref{eq:imp}
using the formalism developed in Chap.~3 of Ref.~\onlinecite{zoller}. We start
by considering the Hamiltonian of a transmon coupled linearly to a linear
environment that without loss of generality can be assumed to be a collection
of harmonic oscillators. The Hamiltonian of the total system can be taken as
\begin{equation}\label{eq:HGZ}
  \hat H= \hat H_{q}(\hat Q, \hat \Phi)+ \tfrac{1}{2} \sum_n \Bigl[(\hat
  p_n-\kappa_n \hat \Phi)^2 +\omega_n^2 \hat q_n^2 \Bigr],
\end{equation}
with the system (qubit) Hamiltonian
\begin{equation}
  \hat H_{q}= \frac{\hat Q^2}{2 C} -E_J \cos \biggl(\frac{2
\pi}{\Phi_0} \hat \Phi \biggr);
\end{equation}
note that this definition is equivalent to Eq.~\eqref{eq:hq} with $\hat
\varphi=2 \pi \hat \Phi/\Phi_0$,  $\Phi_0=h/2e$, and $I_c= 2\pi E_J/\Phi_0$.
We further impose the following commutation relations for the bath
\begin{subequations}
\begin{equation}
[\hat p_n, \hat p_m]=[\hat q_n, \hat q_m]=0,
\end{equation}
\begin{equation}
[\hat q_n, \hat p_m]= i \hbar \delta_{nm},
\end{equation}
\end{subequations}
and for the system
\begin{equation}
[\hat \Phi, \hat Q]=i \hbar,
\end{equation}
while any system operators commutes with any bath operator. These commutation
relations will hold true between operators in the Heisenberg picture at the
same time. Following Ref.~\onlinecite{zoller}, we can show that from the
Hamiltonian Eq.~\eqref{eq:HGZ}, our Eq.~\eqref{eq:ccl} follows, with the admittance in the time domain identified as
\begin{equation}
Y(t)= \begin{cases}
0, \quad t<0, \\
\sum_n \kappa_n^2 \cos(\omega_n t), \quad t \ge 0,
\end{cases}
\end{equation}
which is manifestly causal, and the noise current term
\begin{equation}
\hat \imath(t)= i \sum_n \kappa_n \sqrt{\frac{\hbar \omega_n}{2}}
\hat a_n^{\dagger}(t_0)+\text{H.c.}
\end{equation}
with the annihilation operator of the $n$-th harmonic oscillator
$\hat a_n=(\omega_n \hat q_n+i \hat p_n)/\sqrt{2 \hbar \omega_n}$
and $\hat a_n^{\dagger}$ its Hermitian conjugate.

After these identifications we perform the two level approximation  by
projecting the system Hamiltonian and the operator $\Phi$ onto the first two
levels. This leads to the following substitutions 
\begin{align}
  \hat H_{q}(\hat \Phi, \hat Q) &\mapsto \tfrac12 \hbar \omega_q  
  \hat \sigma^z,&
  \hat \Phi &\mapsto \sqrt{\tfrac12 \hbar Z_q} \;\hat \sigma^x.
\end{align}
in the Hamiltonian Eq.~\eqref{eq:HGZ}.
We then obtain the spin-boson Hamiltonian\cite{leppakangas:18}
\begin{equation}\label{eq:HGZQ}
  \hat H= \tfrac12 \hbar \omega_q \hat \sigma^z + \tfrac{1}{2} \sum_n (
  \hat p_n^2+\omega_n^2 \hat q_n^2 )-\sqrt{\frac{\hbar Z_q}{2} } \hat \sigma^x
  \sum_{n}\kappa_n \hat p_n,
\end{equation}
where we have neglected constant terms. The Heisenberg equation of motion for
a generic qubit operator $\hat A_q$ under the Hamiltonian Eq.~\eqref{eq:HGZQ}
reads
\begin{multline}\label{eq:sysOpEOM}
  \frac{d \hat A_q(t)}{dt}=\tfrac{i}2   \omega_q [\hat \sigma^z(t), \hat A_q (t)]\\ 
-i\sqrt{\frac{ Z_q}{2 \hbar}}[\hat \sigma^x(t),
\hat A_q (t)]\sum_n \kappa_n \hat p_n(t),
\end{multline}
where we have used the fact that the operator $\sum_n \kappa_n \hat p_n(t)$
commutes with any system operator at the same time. Following
Ref.~\onlinecite{zoller}, we also obtain
\begin{equation}
\sum_n \kappa_n \hat p_n(t)= \hat \imath (t)
-\sqrt{\frac{\hbar Z_q}{2}} \int_{t_0}^t \!\!dt'\frac{d Y(t-t')}{d t'}
\hat \sigma^x(t');
\end{equation}
in what follows, we will neglect the transient behavior and  let $t_0 \to
-\infty$. We want to obtain the Heisenberg equation of motion for the operator
$\hat \sigma^-(t)$. From Eq.~\eqref{eq:sysOpEOM}, we readily obtain
\begin{multline}
  \partial_t \hat\sigma^-(t) +i \omega_q \hat \sigma^-(t)-\tfrac{i}2 Z_q
\sigma^z(t)\int \!dt' \frac{d Y(t-t')}{d t'} \hat \sigma^x(t')\\
= 
-i \sqrt{\frac{Z_q}{2 \hbar}} \hat \sigma^z(t) \hat \imath(t).
\end{multline}
Going over to a rotating frame with $\hat \sigma^- \mapsto \hat \sigma^- e^{-i
\omega_q t}$, like in Eq.~\eqref{eq:rw_ansatz}, yields
\begin{multline}
  e^{-i \omega_q t}\hat \sigma^z (t)\partial_t \hat\sigma^-(t) - \tfrac{i}2
Z_q\int \! dt'  \frac{d Y(t-t')}{d t'}\\
\times[e^{-i \omega_q t'}
\hat \sigma^-(t')
+\text{H.c.}]  = -i \sqrt{\frac{Z_q}{2 \hbar}}  \hat
\imath(t).
\end{multline}
Taking the Fourier transform, we obtain
\begin{multline}
-(\hat \sigma^z \partial_t \hat\sigma^-)_{\eta}\!+\!\tfrac12 \omega Z_q
Y_{\omega}\bigl(\hat \sigma_\eta^- + \hat\sigma_{\omega+\omega_q}^+ \bigr)\!=\!i
\sqrt{\frac{Z_q}{2 \hbar}} \hat \imath_{\omega},
\end{multline}
Now neglecting the term proportional to $\hat\sigma_{\omega+\omega_q}^+ $
consistently with the approximation described in the text we finally obtain
Eq.~\eqref{eq:imp}.

The result of the Appendix can be summarized as follows: in the two-level
approximation, the current $\hat I_q =\hbar C \ddot{\hat \varphi}/2e + I_c
\sin\hat\varphi$ that is flowing through the qubit assumes the form
\begin{equation}
  \hat I_q(t) = i \sqrt{\frac{2 \hbar}{Z_q}} \;\hat \sigma^z(t) \partial_t \hat
  \sigma^-(t) + \text{H.c.}
\end{equation}
with $\sigma^-$ in the rotating frame. This result can be used in the equation
of motion \eqref{eq:ccl} to project it onto the qubit subspace.

\section{Admittance of a multi-mode stripline resonator}\label{app:admit}

The admittance of a transmission line or stripline which is shunted by a load
with admittance $Y_L$ is given by\cite{pozar}
\begin{equation}\label{eq:full_admittance}
  Y_\text{tl}=Z_0^{-1} \frac{Z_0 Y_L-i \tan(\pi
  \omega/\omega_0)}{1-i Z_0 Y_L \tan(\pi \omega/\omega_0)}
\end{equation}
with $Z_0$ the characteristic impedance of the transmission line and $\omega_0$
the fundamental frequency. We are interested in the situation where the
transmission line forms a good (multi-mode) resonator. In this case, we have
that $|Y_L| Z_0 \ll 1$ and we can use the expansion
\begin{equation}\label{eq:full_admittance2}
  Y_\text{tl}=-i Z_0^{-1}  \tan(\pi
  \omega/\omega_0) + Y_L
\end{equation}
where we have made use of the fact that the second term is only relevant when
$\tan(\pi \omega/\omega_0) \ll 1$. We also can use the alternative expression
$Y_\text{tl}=-i Z_0^{-1}  \tan[\pi
  ( \omega + i \kappa_\omega/2)/\omega_0]$ with 
  \begin{equation}\label{eq:kappao}
  \kappa_\omega = \frac{2 \omega_0 Z_0
Y_L}{\pi}
\end{equation}
valid to the same order. Here, we have made explicit that in principle
$\kappa_\omega$ depends on frequency via $Y_L$. However, as it is only
important to accurately describe  the admittance close to qubit frequency, we
can approximately set $\omega = \omega_q$ in Eq.~\eqref{eq:kappao}.  For
concreteness, we require $C_L \omega_q R \gg 1$ in order to obtain
Eq.~\eqref{eq:mm} with
\begin{align}
  \kappa = \frac{2 \omega_0 Z_0}{\pi R} \qquad \text{and} \qquad
   C_r = \frac{\pi}{2 \omega_0 Z_0}.
\end{align}
For typical experiments, this regime is not obtained. As a result the decay
rate of the modes depends on the modenumber \cite{houck:08}. Our approach also
works in this regime, however the results are not so nice as the admittance is
not given by \eqref{eq:admit} but rather $\kappa$ becomes frequency dependent.

\section{Integrating out the resonator}\label{app:int_out}

In this appendix, we would like to show that the equations \eqref{eq:imp_r}
and \eqref{eq:imp_r2} are equivalent to \eqref{eq:imp} after the resonator
mode $\hat a_\eta$ has been integrated out and thus the node $R$ eliminated.
Solving \eqref{eq:imp_r2}  for $\hat a_\eta$ yields
\begin{equation}
  \hat a_\eta = \frac{\tilde Y_\omega \sqrt{\frac{C_r}{C}}\, \hat \sigma^-_\eta +i \sqrt{\frac{2
  C_r}{\hbar \omega_q}}( \hat{
  \imath}_{r,\omega} - \hat{\tilde
\imath}_\omega )  }{Y_{r,\omega} + \tilde Y_{\omega}} .
\end{equation}
Plugging this expression into \eqref{eq:imp_r}, we obtain \eqref{eq:imp}
with $Y^{-1} = Y_{r}^{-1} + \tilde Y^{-1}$ and
\begin{equation}
  \hat{\imath}_\omega  = \frac{\tilde Y_\omega \hat 
    \imath_{r,\omega} + Y_{r,\omega}
  \hat{\tilde \imath}}{Y_{r,\omega} + \tilde Y_{\omega}}.
\end{equation}
Due to the relation
\begin{equation}
  \operatorname{Re}(Y_\omega)  = \frac{|\tilde Y_\omega|^2
    \operatorname{Re}(Y_{r,\omega} )  + |\tilde Y_{r,\omega}|^2
  \operatorname{Re}(\tilde Y_{\omega} ) }{|Y_{r,\omega} + \tilde Y_{\omega}|^2}.
\end{equation}
and the fact that $ \hat \imath_{r,\omega}$ and $\hat{\tilde \imath}$ are
distributed according to Eqs.~\eqref{eq:comm} and \eqref{eq:noise} with $Y$
replaced by $Y_r$ and $\tilde Y_{\omega}$, it can be shown that $\hat\imath$
has  the correct commutation relation and expectation value.

\section{Rediagonalization approach}\label{app:rediag}

In this section, we present the rediagonalization approach of
Ref.~\onlinecite{gely:17} for the stripline resonator. However, for
convenience, we diagonalize the system already on the Lagrangian level and not
on the Hamiltonian as in Ref.~\onlinecite{gely:17}.

The Lagrangian of the qubit coupled to a stripline resonator (with $\kappa=0$)
is given by
\begin{align}\label{eq:jc-lagrangian}
  \mathcal{L}&=\frac{C+C_c}{2} \dot{\varphi}^2+ \frac{\hbar I_c}{2e}
  \cos(\varphi)-C_c\dot{\varphi}\sum_{\mu=0}^\infty
  \dot{\phi}_\mu + \mathcal{L}_r, \\
  \mathcal{L}_r&= C_r \dot \phi_0^2+\sum_{\mu=1}^\infty
  \left( \frac{C_r \dot{\phi}_\mu^2}{2} -\frac{\phi_\mu^2}{2
  L_\mu}\right)+\frac{C_c}{2}\left(\sum_{\mu=0}^\infty\dot{\phi}_\mu\right)^2\,,
\nonumber
\end{align}
where $L_\mu =(\mu^2 C_r \omega_0^2)^{-1}$ for $\mu\in\mathbb{N}$.  That
this is the correct description of the environment, can  be seen by the
expansion
\begin{align}
  Y^{-1}_\text{mm} &=
  \frac{\pi i}{2 C_r \omega_0} \cot(\pi\omega/\omega_0) \nonumber\\
  &=   (- i  C_r \omega)^{-1} \left[\frac12 + \sum_{\mu=1}^\infty \left(1-
    \frac{\mu^2 \omega_0^2}{\omega^2}\right)^{-1}  \right]
\end{align}
which in circuit terms corresponds to a capacitance in series with an infinite
ladder of LC-resonators.

The method proceeds by finding the eigenmodes of the resonator Lagrangian
$\mathcal{L}_r$ including the coupling capacitance $C_c$. In particular, we
would like to find eigensolutions to the Euler-Langrange equations 
\begin{align}
  (C_r + C_c) \ddot\phi_\mu + L_\mu^{-1} \phi_\mu + C_c \sum_{\lambda\neq
  \mu} \ddot \phi_\lambda =0
\end{align}
with $\phi_\mu(t) = e^{-i \nu t} v_\mu$. This corresponds to the generalized
eigenvalues problem $ A\bm v = \nu^2 B \bm v$ where $A$ is diagonal with
$A_{\mu\lambda} = L_\mu^{-1} \delta_{\mu\lambda}$ and  $B_{\mu\lambda} = C_r
\delta_{\mu\lambda} + C_c + C_r \delta_{\mu0}\delta_{\lambda0}$.

From the general theory of symmetric generalized eigenvalue problems with
positive definite matrices, it is known that the eigenvalues $\nu_n^2$ are
positive (thus we can choose $\nu_n\geq 0$) and that the eigenvectors can be
normalized such that
\begin{equation}
  {\bm v}_n \cdot B {\bm v}_m = C_r \delta_{nm}
\end{equation}
Thus, introducing the new modes $\psi_n(t)$ with
\begin{equation}
 \bm \phi(t) = \sum_{n=0}^\infty \psi_n(t) \bm v_n
\end{equation}
the Lagrangian assumes the diagonal form
\begin{multline}
  \mathcal{L} = \frac{C+ C_c}2 \dot \varphi^2 + \frac{\hbar I_c}{2e}
  \cos(\varphi) - \dot \varphi \sum_{n=0}^\infty C_n \dot\psi_n \\
  + \frac{C_r}2 \sum_{n=0}^\infty (\dot \psi_n^2 - \nu_n^2 \psi_n^2)
\end{multline}
with the coupling capacitance
\begin{equation}
  C_n = C_c \sum_{\mu=0}^\infty (\bm v_n)_\mu
\end{equation}
to the $n$-eigenmode. 

In order to define coupling strength $g_n$, the authors of
Ref.~\onlinecite{gely:17} propose to treat the case of a qubit in
the transmon regime. In this case, the Josephson junction effectively acts as
an inductance.  The coupling strength to the $n$-th mode can be
identified with
\begin{equation}
  g_n  = \frac{ C_n \omega_q }{2\sqrt{(C+C_c) C_r}} ,
\end{equation}
see Eq.~\eqref{eq:g}.
Note that crucially, the strength now depends on $n$ as $C_n$ is not simply
$C_c$. The equations of motion are given by
\begin{align}
  (C+C_c) (\ddot \varphi + \omega_q^2 \varphi) &= \sum_{n=0}^\infty C_n
  \ddot\psi_n, \label{eq:qu_eq}\\
  C_r ( \ddot \psi_n + \nu_n^2 \psi_n) &= C_n\ddot \varphi.
\end{align}

Going over to frequency space and solving for the mode $\psi_{n}$, we obtain
\begin{align}\label{eq:phi_sol}
  \psi_{n,\omega} = \frac{C_n \omega^2 \varphi_\omega}{C_r(\omega^2-\nu_n^2
  )}
  \,.
\end{align}
The qubit equation Eq.~\eqref{eq:qu_eq} involves only frequencies close to the
qubit frequency.  Assuming that all the frequencies $\nu_n$ are sufficiently
detuned from the qubit frequency $\omega_q$, we can replace $\omega\mapsto
\omega_q$ in Eq.~\eqref{eq:phi_sol}. Plugging this into the
Eq.~\eqref{eq:qu_eq}, we obtain
\begin{align}
  0 &= \ddot \varphi +\omega_q^2  \left(1 +  \sum_{n=0}^\infty \frac{4
   g^2_n}{\omega_q^2-\nu_n^2}
 \right)
\varphi \nonumber\\
    &\approx \ddot \varphi + \left(\omega_q +  \sum_{n=0}^\infty \frac{2
   \omega_q g^2_n}{\omega_q^2-\nu_n^2}
 \right)^2
\varphi ,
\end{align}
i.e, the shift due to the different modes is independent as announced by
Ref.~\onlinecite{gely:17}.

Numerically, the procedure is as follows. The number of modes in
Eq.~\eqref{eq:jc-lagrangian} is made finite with $\mu \leq N$. Then the
generalized eigenvalue problem is solved on a computer. The eigenenergies
$\nu_n$ as well as the coupling constants $g_n$ are calculated. As the
different eigenmodes act independently, the Lamb shift is approximately given
by the sum of the Lamb shifts of the individual modes, i.e.,
\begin{equation}
\delta \omega_q = \sum_{n=0}^\infty  \frac{2 \omega_q g_n^2 }{\omega_q^2 -
\nu_n^2} = \sum_{n=0}^\infty 
  \frac{g_n^2}{\Delta_n(1-\Delta_n/2\omega_q)}
\end{equation}
with $\Delta_n = \omega_q -\nu_n$.
Note that due to the fact that some of the modes are highly detuned, we could not
use the approximation
\begin{equation}
  \frac{2 \omega_q} {\omega_q^2 - \nu_n^2} = \frac{1}{\Delta_n}
\end{equation}
that is commonly used in the dispersive regime with $\Delta_n \ll \omega_q$.

\end{document}